\documentclass[twocolumn,prd,superscriptaddress,nofootinbib]{revtex4-1}

\usepackage{amsfonts,amsmath,amssymb,mathrsfs}
\usepackage{hyperref}
\usepackage{color}
\usepackage{graphicx} 
\usepackage{dcolumn}  
\usepackage{bm}    
\usepackage[english]{babel}
\usepackage[bottom]{footmisc}
\usepackage{dblfnote}
\usepackage{mathtools}
\usepackage{amsmath}
\usepackage{physics}
\DFNalwaysdouble 

\def\a{\alpha}

\def\r{\rho}
\def\s{\sigma}
\def\t{\tau}
\def\m{\mu}
\def\n{\nu}
\def\k{\kappa}
\def\th{\theta}
\def\g{\gamma}\def\G{\Gamma}
\def\L{t}\def\l{V}
\def\D{\Delta}
\def\la{\langle}
\def\ra{\rangle}
\def\o{\omega}\def\O{\Omega}
\def\d{\delta}
\def\p{\partial}

\def\oxthree{{\cal O}(x^3) }

\def\half{\textstyle{\frac{1}{2}}}

\def\bdoc{\begin{document}}
\def\edoc{\end{document}}
\def\bea{\begin{equation}}
\def\eea{\end{equation}}

\def\beq{\begin{eqnarray}}
\def\eeq{\end{eqnarray}}
\def\be{\begin{eqnarray}}
\def\ee{\end{eqnarray}}
\def\ben{\begin{enumerate}}
\def\een{\end{enumerate}}
\def\la{\langle}
\def\ra{\rangle}
\def\a{\alpha}
\def\g{\gamma}\def\G{\Gamma}
\def\d{\delta}\def\D{\Delta}
\def\e{\epsilon}
\def\z{\zeta}

\def\th{\theta}
\def\k{\kappa}
\def\l{t}
\def\m{\mu}
\def\n{\nu}
\def\o{\omega}
\def\p{\pi}
\def\r{\rho}
\def\s{\sigma}
\def\t{\tau}
\def\L{{\cal L}}
\def\S{\Sigma }
\def\gsim{\; \raisebox{-.8ex}{$\stackrel{\textstyle >}{\sim}$}\;}
\def\lsim{\; \raisebox{-.8ex}{$\stackrel{\textstyle <}{\sim}$}\;}
\def\gtrsim{\gsim}
\def\lessim{\lsim}
\def\loc{{\rm local}}
\def\vm{v_{\rm max}}
\def\bh{\bar{h}}
\def\del{\partial}
\def\nab{\nabla}
\def\half{{\textstyle{\frac{1}{2}}}}
\def\fourth{{\textstyle{\frac{1}{4}}}}

\def\bD{{\bf D}}
\def\bE{{\bf E}}
\def\bF{{\bf F}}
\def\bB{{\bf B}}
\def\bP{{\bf P}}
\def\bV{{\bf v}}
\def\bv{{\bf v}}
\def\bx{{\bf x}}
\def\by{{\bf y}}
\def\bz{{\bf z}}
\def\ba{{\bf a}}
\def\bd{{\bf d}}
\def\bs{{\bf s}}
\def\bn{{\bf n}}
\def\bp{{\bf p}}

\def\O{\Omega}

\def\br{{\bf r}}
\def\bnab{{\bf \nab}}

\def\tE{\tilde{E}}
\def\tL{\tilde{L}}
\def\Horava{Ho\v{r}ava }

\def\oxtwo{\mathscr{O}\left(x^2\right)}
\def\oxthree{\mathscr{O}\left(x^3\right)}
\def\oxfour{\mathscr{O}\left(x^4\right)}
\def\oxfive{\mathscr{O}\left(x^5\right)}
\def\LL{\text{Lanczos-Lovelock}}

\def\ph{\phantom}

\begin{document}
\title{
Light rings of stationary spacetimes}
\author{Rajes Ghosh}
\email{rajes.ghosh@iitgn.ac.in }
\affiliation{Indian Institute of Technology, Gandhinagar, Gujarat 382355, India.}
\author{Sudipta Sarkar}
\email{sudiptas@iitgn.ac.in}
\affiliation{Indian Institute of Technology, Gandhinagar, Gujarat 382355, India.}

\begin{abstract}
We present a novel theorem regarding light rings in a stationary spacetime with an ergoregion. We prove that any stationary, axisymmetric, and asymptotically flat spacetime in $1+3$ dimensions with an ergoregion must have at least one light ring outside the ergoregion. A possible extension of the proof for asymptotically de-Sitter and anti-de-Sitter spherically symmetric black holes is also discussed. 
 \end{abstract}

\maketitle

\section{Introduction}
Astrophysical compact objects are the natural laboratories to test different theories of gravity. Various observational probes of these objects, such as gravitational waves, quasi-normal modes, shadows etc., may reveal hitherto unknown aspects of the strong gravitational field. All these observational tools crucially hinge upon two important ingredients associated with the spacetime outside a compact object: the event horizon and the light ring (LR). The event horizon is a null uni-directional hypersurface that serves as a causal boundary. The existence of an event horizon means the compact object is actually a black hole (BH). In contrast, a light ring is a null orbit around a compact object which causes an extreme deflection of light-rays. The observation of the dark shadow due to gravitational light bending and photon capture by the supermassive black hole candidate in the centre of the giant elliptical galaxy M87 unequivocally confirms the existence of light rings around the central compact object \cite{Akiyama:2019cqa}. However, the existence of LRs does not uniquely determine the nature of the central object. There may exist horizonless ultra-compact objects (UCOs) which also support light rings.\\

All exact solutions describing the spacetime outside compact objects assume a high degree of symmetry, which may not be true for astrophysical systems. The lack of symmetry may induce significant changes in various properties of the spacetime and can cause substantial modifications in the observational signatures. At this stage, we can ask a crucial question: Do compact objects in equilibrium always support light rings? In general relativity (GR), this question bears an affirmative answer for the unique stationary black hole solutions of electro-vacuum Einstein's equations known as the Kerr-Newman spacetime. This may not be the general feature for any arbitrary theory of gravity. However, in paper \cite{Cunha2}, the authors proved a robust theorem: ``A stationary, axisymmetric, asymptotically flat, $1+3$ dimensional BH spacetime with a non-extremal, topologically spherical Killing horizon, admits at least one standard LR outside the horizon for each rotation sense''. The authors used novel topological arguments to prove this remarkable result. The validity of the theorem is theory-independent, though their arguments depend crucially on the spacetime dimensions and the topology of the Killing horizon of the black hole. A similar proof \cite{Wei} also exists for black holes in asymptotically flat static de-Sitter and anti-de-Sitter cases. This proof is also topological and valid in $D=4$ only. \\

Another recent work \cite{Cunha1} (also see \cite{Guo:2020qwk}) has shown that if we assume that there is one light ring outside a horizonless compact object, there must be a second one. In fact, any such UCOs originating from gravitational collapse necessarily have an even number of light rings. Note that the spacetime outside a spinning charged star need not be Kerr-Newman\\

In this letter, we prove a stronger theorem: \textit{In any stationary, axisymmetric, and asymptotically flat spacetime in $1+3$ dimensions with an ergoregion must have at least one light ring outside the ergoregion}. We show that the behaviour of various metric components at the ergoregion and their asymptotic fall-off (or asymptotic growth) conditions at large distances are enough to prove the claim. Our method is entirely algebraic, and therefore, topological concepts such as fixed points of a map are not required. We also discuss some possible generalization of our algebraic method to asymptotically de-Sitter(dS) and anti-de-Sitter (AdS) spacetimes by considering simple spherically symmetric illustrations.\\

Our theorem shows that given the existence of an ergoregion, there must be at least one light ring outside. This does not require the existence of an event horizon. Nevertheless, if we assume the presence of a stationary event horizon inside the ergoregion, it implies the result proved in \cite{Cunha2}. For the case of rotating UCOs with ergoregion, our theorem justifies the assumption used in \cite{Cunha1} to show that light rings occur in pairs. Some of these light rings are also shown to be stable. \\

It was suggested that the presence of an ergoregion in a stationary, asymptotically flat spacetime without any horizon makes the configuration unstable under linear perturbations \cite{Friedman}. Such instability is phrased in literature as `Ergoregion instability'. However, it is possible to construct models of horizonless rotating compact objects with ergoregion, which are stable \cite{Chirenti:2008pf}. On the other hand, for the case of static UCOs with stable light rings, spacetime instability occurs under nonlinear perturbations which may cause fragmentation of the ultracompact stars and the destruction of the light ring \cite{Cardoso:2014sna}. Though there is no rigorous proof, a similar instability is also expected to develop in the stationary case as well. If this is indeed the case, our theorem, along with \cite{Cunha2}, suggests that even if a rotating UCO is stable under `Ergoregion instability', it would nevertheless suffer from instability at a nonlinear level due to the presence of a stable light ring. This significantly strengthens the argument in favour of the so-called ``black hole hypothesis", which claims that the objects with a light ring are black holes.

\section{Light rings of stationary compact objects}\label{stat}
Let us start with the geometric setup describing the spacetime of a compact object. In 4-dimensions, if the central object is a stationary black hole, then it is a well-known result that the spacetime must also be axisymmetric \cite{Hawking}. In other words, for a rotating black hole spacetime in $D=4$, the existence of a timelike Killing vector $\partial_t$ implies the existence of a spacelike isometry $\partial_{\phi}$. However, the proof of this rigidity theorem requires Einstein's field equations. Thus, in a general theory of gravity, such uniqueness may not hold true. In addition, we are considering the case when the central object may not have a horizon and therefore, the rigidity theorem does not hold true even in GR. Nevertheless, for simplicity, we assume the existence of both the Killing vectors: $\partial_t$ and $\partial_{\phi}$, in our setup. This assumption also leads to a well-defined notion of light rings. We also use another assumption that the metric is invariant under the simultaneous reflections $t \to -t$ and $\phi \to -\phi$. Then, our setup is identical to the case considered in \cite{Cunha1, Cunha2}. \\

Using these assumptions, the most general metric for an asymptotically flat stationary spacetime having two isometries $\partial_t$ and $\partial_{\phi}$ in $D=4$ is given by,
\bea \label{statm}
ds^2=g_{tt}dt^2+g_{rr}dr^2+g_{\theta \theta}d\theta^2+g_{\phi \phi} d\phi^2+2g_{t \phi} dt d\phi\ .
\eea
Note that the existence of the aforesaid isometries demand that all metric components should be independent of $t$ and $\phi$ coordinates. If the central object is a black hole, the location of the event horizons are the positive roots of the equation $g_{rr}=0$. Moreover, outside the rotating central object, we have $\ g_{rr} > 0,\ g_{\theta \theta} >0,\ \text{and}\ g_{\phi \phi} > 0\ (\text{away from the axis})$ \cite{Cunha1}. Note also that the asymptotic flatness requires the following behaviour of various metric components at large r: $ g_{t t} \to -1+C/r + {\cal O}(1/r^2)\ ,\ g_{t \phi} \sim \pm\ r^{-1} \text{sin}^2(\theta)\ \text{and},\ g_{\phi \phi} \sim \ r^{2} \text{sin}^2(\theta)\ .$\\

The location of the ergoregion is associated with positive roots of the equation $g_{tt}=0$. The ergosphere is a Killing horizon $\mathcal{H}$ which assumed to have topologically spherical cross sections, which is also used in \cite{Cunha2}. Let $r=r_e$ denotes the position of the outermost ergoregion. At this stage, we assume that the ergoregion is a non-extremal Killing horizon of the Killing vector $\partial_t$. Then, $g_{tt}$ is positive just inside the outermost ergoregion, zero on the ergoregion, and negative outside of it. Thus, we must have $-g_{tt}'(r_e)>0$. Now, to find out the location ($r=r_l$) of the LRs, we construct the following functions \cite{Cunha1},
\bea \label{statf}
H_{\pm}(r,\theta) = \frac{-g_{t \phi} \pm \sqrt{\Delta}}{g_{\phi \phi}},
\eea
where $\Delta=g_{t \phi}^2-g_{tt}g_{\phi \phi}$ and $\Delta > 0$ outside the central object. The $\pm$ signs are there to distinguish the two opposite sense of rotations. Next, we define the light ring exactly as in \cite{Cunha1}: A LR is a null geodesic with a tangent vector field that is always a linear combination of the Killing vectors
$\partial_t$ and $\partial_\phi$ only. This implies following condition on the photon momentum, $ p_\mu = \dot{p}_\mu = 0$ where $\mu = r, \theta$. This condition can also be reformulated using an effective potential $V$, such that at the location of the light ring, $V = \nabla V = 0$. In \cite{Cunha1}, it is shown that this condition on the effective potential translates into conditions on the functions $H_{\pm}$; a light ring must be the critical points of the function. At the location of the LRs, we must have: Either $\partial_{\mu} H_{+}=0$, or $\partial_{\mu} H_{-}=0$, or both for the static case. To prove our theorem, we have to show there exists at least one root $\{r=r_l,\theta=\theta_l\}$ of the equation $\partial_{\mu} H_{\pm}=0$ in the region $r_e \leq r < \infty$ and $0 < \theta < \pi$.\\

Our first step towards proving the theorem will be to show that for all values of $r > r_e$, there exists at least one solution $\theta_0 \in (0,\pi)$ of the equation $\partial_{\theta} H_{\pm}=0$. For this purpose, observe that for any fixed value of $r \in (r_e,\infty)$, the effective potentials behave as \cite{Cunha2}: $H_{\pm} \sim \pm 1/\rho$ near the axis. Here, we have introduced a local coordinate $\rho = \sqrt{g_{\phi \phi}}$ near the axis with the property that $\partial_\theta \rho$ is positive (negative) as $\theta \to 0$ ($\theta \to \pi$). Therefore, we have \cite{Cunha2},
\bea \label{beh}
\partial_{\theta} H_{\pm} \sim \mp \frac{\partial_{\theta} \rho}{\rho^{2}} \sim\left\{\begin{array}{ll}
\mp \infty, & \text{as}\ \theta \rightarrow 0 \\
\pm \infty, & \text{as}\ \theta \rightarrow \pi\ .
\end{array}\right.
\eea
 It immediately suggests that $\partial_\theta H_{\pm}$ must have at least one zero between $(0,\pi)$ at all values of $r > r_e$. As r varies, we get a trajectory of solutions $\theta_0(r)$ of the equation $\partial_{\theta} H_{\pm}=0$.\\

On the other hand, using Eq.(\ref{statf}), we can express $\partial_r H_{\pm}(r,\theta)$ in the following suggestive way:
 \bea \label{statlr}
 \partial_r H_{\pm}(r,\theta)=\pm \frac{1}{2 \sqrt{\Delta}\ g_{\phi \phi} }\left[L(r,\theta)-R(r,\theta)\right],
 \eea
where $L(r,\theta)= -g_{tt}' g_{\phi \phi},$ and $R(r, \theta)= -g_{tt} g_{\phi \phi}'\ \pm\ \left(2/g_{\phi \phi}\right) \left(\sqrt{\Delta} \mp g_{t \phi} \right)\left(g_{t \phi}'g_{\phi \phi} - g_{\phi \phi}' g_{t \phi}\right)$. At the location of light rings, we have equality between these two functions. Now, we want to study their behaviour in the range $r_e \leq r < \infty$, where they satisfy the following properties--\\
\\(i) At the outermost ergoregion $r=r_e$, the function $R(r=r_e)=0$. It is due to the fact that $g_{t t}(r_e)=0$ and $\sqrt{\Delta(r_e)} = \pm g_{t \phi}(r_e)$. Furthermore, we have $R(r) \sim r$ as $r \to \infty$.\\
\\(ii) On the other hand, the function $L(r,\theta)>0$ at $r=r_e$ and approaches to unity asymptotically, i.e., $L(r) \to 1$ at large r.\\
\\These two properties make it clear that the functions $L(r)$ and $R(r)$ must have at least one intersection $r=r_0(\theta)$ in the region $r_e < r < \infty$ for all values of $\theta \in (0, \pi)$.\\
\\Now, we have two functions $r=r_0(\theta)$ and $\theta = \theta_0(r)$ satisfying two separate equations $\partial_r H_\pm=0$ and $\partial_\theta H_\pm =0$, respectively. Since for all $\theta_0 \in (0,\pi)$, there exists a point on the curve $r=r_0(\theta_0)$ so that $r_e < r_0 < \infty$ and vice versa, it is obvious that these two curves must intersect at least at a point $(r_l,\theta_l)$ where both equations are satisfied. This ensures the existence of at least one light ring outside the ergoregion and thereby completes the proof of the theorem.\\
\\Though the theorem is proved at $D=4$ spacetime dimensions, the technique used here begs a natural extension to higher dimensions. Our proof hinges very crucially on the asymptotic behaviour of various metric components and therefore, this proof breaks down for spacetimes which are not asymptotically flat.\\

We now illustrate an application of our method for the static and spherically symmetric spacetimes. In fact, for such cases, our method can also be extended to any spacetime dimensions greater than three.

\section{Light rings of spherically symmetric black holes}\label{sph}
 The metric of a static and spherically symmetric black hole spacetime is given by:
\bea \label{sphm}
ds^2 = - f(r)\, dt^2 + \frac{1}{g(r)}dr^2 + h(r) d\Omega_{(D-2)}^2,
\eea
where, $d\Omega_{(D-2)}^2$ is the metric on a unit (D-2)-sphere and asymptotic flatness requires at large $r$,
\begin{align} \label{sphf}
f(r) \to 1 - \frac{ C}{r^{D-3}} + {\cal O}(1/r^{D-2})\ ; \nonumber\\
\text{and,}\ h(r) \sim r^2 \ .
\end{align}
Where $C$ is related to the ADM mass of the spacetime and the positive mass theorem implies $C>0$. Let $r=r_h$ represents the position of the \textit{outermost root} of the equation $f(r)=0$. We shall also assume the root at $r=r_h$ is non-degenerate. Then, the function $f(r)$ is negative just inside and positive just outside $r=r_h$. In other words, we have the inequality: $f'(r_h) > 0$.\\

Using the null geodesic equation, we find the following condition for LRs \cite{Wei, Mishra:2019trb}: $h(r_l) \, f'(r_l) = f(r_l)h'(r_l)$. This is a special case of the stationary spacetime described in the previous section. It is easy to show that for a Schwarzschild black hole of mass $m$, this implies $r_l = 3 m$.\\

We want to prove that for a general spherically symmetric spacetime, the equation of the light ring always admit at least one solution outside the outermost root of the function $f(r)$. For this purpose, as in the previous section, we consider two functions: $L(r)= h(r) f'(r)$, and $R(r) = f(r)h'(r)$. We want to study their behaviour in the range $r_h \leq r < \infty$, where they satisfy the following properties--\\
\\(i) At $r=r_h$, the function $R(r)$ vanishes. Furthermore, $R(r) \sim r$ at large r.\\
\\(ii) On the other hand, the function $L(r)>0$ at $r=r_h$ and at large r, its the fall-off condition is given by $L(r) \sim r^{-(D-4)} $.\\
\\Using these two properties, it is obvious that the functions $L(r)$ and $R(r)$ have at least one intersection ($r=r_l$) in the region $r_h \leq r < \infty$ for any $D\geq 4$. Thus, there must be at least one solution of the equation for LRs outside the outermost root of the function $f(r)$. Note that for the metric \ref{sphm}, the event horizon is a null hypersurface which is determined from the equation $g(r) = 0$. Whereas the zeroes of $f(r)$ imply the vanishing of the norm of the timelike Killing vector $\partial_t$. Therefore, the outermost root of the equation $f(r) = 0$ is the surface of static limit.\\

Interestingly, one can show that the finiteness of the Ricci scalar $R$ at the surface where $f(r) =0$ requires the function $g(r)$ to vanish as well at the same location. However, the opposite statement need not be true. The function $g(r)$ may have zeroes outside the static limit where $f(r)=0$, i.e., the static-limit surface may not coincide with the outermost event horizon. Nevertheless, if we assume the validity of Hawking's strong rigidity theorem, the event horizon where $g(r) = 0$ must also be a Killing horizon, where the norm of a timelike Killing vector vanishes. Since there is only one timelike Killing vector for this spacetime, the surface $g(r) = 0$ must coincide with the Killing horizon at $f(r) = 0$. Then, our result implies that there must exist at least one LR outside the outermost event horizon. This is exactly the case with Schwarzschild black holes where $f(r)$ is the same as $g(r)$.\\

Unlike the stationary case, the above proof is valid for any dimensions $D\geq 4$. Though our proof goes through for either sign of $f'(r)$ at large r $\left(\text{i.e.,}\ f'(r) \sim \pm\ r^{-(D-2)}\right)$, positive mass theorem fixes the sign to be positive. Note, however, that the above method breaks down for asymptotically de-Sitter (dS) or anti-de-Sitter (AdS) cases. It is because of the fact that in both of these cases, the functions $L(r)$ and $R(r)$ have the same growth at large r. Let us now discuss such a situation with simple examples of Schwarzschild-dS (AdS) black holes.\\

The line element of a Schwarzschild-dS (AdS) black hole has the same form as given by Eq.(\ref{sphm}), where the functions $f(r)=g(r)=1-2m/r - \Lambda r^2/3 $, and $h(r) = r^2$. The sign of the cosmological constant is positive (negative) for Schwarzschild-dS (AdS) case. At the location of the outermost event horizon ($r=r_h$) we must have $f(r_h)=0$:

\bea
\text{Schwarzschild-dS}: \Lambda r_h^3 -3r_h+6m=0\ ,
\eea
\bea
\text{Schwarzschild-AdS}: |\Lambda| r_h^3 +3r_h-6m=0\ .
\eea
However, using the light ring equation, we get $r_l=3m$, irrespective of the cosmological constant($\Lambda$). Our goal is to check whether the LR lies outside the outermost event horizon, i.e., does the relation $r_h < r_l$ hold true for both Schwarzschild-dS and Schwarzschild-dS spacetimes?\\

To investigate the case with $\Lambda >0$, we consider the function $G(r) = \Lambda r^3 -3r+6m$. This function generates positive outputs at $r=0$ and at large values of r as well. In between these two extreme positions, using Descarte's rule of signs, the function can be shown to have at most two zeros, the larger one being the position of the outer horizon ($r=r_h$). The case where $G(r)$ has no positive real root is, in fact, associated with a naked singularity and does not serve our purpose. Thus, we have to choose the parameters $\{\Lambda,\ m\}$ in such a manner that $G(r)$ has two distinct real roots and this is achieved when the condition $9 \Lambda m^2 > 1$ holds true. Note, when $9 \Lambda m^2 = 1$ is satisfied, we have an extremal black hole with a doubly degenerate horizon at $r=1/\sqrt{\Lambda}$.\\
\\In the case of a non-extremal black hole the condition $9 \Lambda m^2 > 1$ is satisfied and the spacetime has two event horizons. The function $G(r)$ is negative in between two horizons and non-negative elsewhere. Now, the positivity of the functions $G(3m)$ and $G'(3m)$ conclusively suggests that the inequality $3m>r_h$ holds true. This result guarantees that \textit{the LR (at $r_l=3m$) indeed lies outside the outermost event horizon of a non-extremal Schwarzschild-dS black hole}. A similar result can also be established for the case of a Schwarzschild-AdS black hole.

\section{Discussions and conclusion}\label{con}
The existence of light rings is a non-trivial feature of a spacetime that has several important consequences. The presence of a horizon, as proved in \cite{Cunha2}, necessarily implies the existence of a light ring. However, the opposite statement need not be true; the presence of light rings does not necessarily mean the existence of a horizon in a spacetime. For example, there exist ultracompact objects which support light rings without having any horizon. Nevertheless, the result in \cite{Cunha1} showed that if a horizonless compact object is assumed to have one light ring, it must have another one, as light rings for such ultracompact objects occur in pairs. Moreover, at least for spherically symmetric case, some of the light rings must be stable. A stable light ring outside a UCO is believed to have nonlinear instability \cite{Cardoso:2014sna} and consequently, such instability, if it exists, provides an argument in favour of the black hole hypothesis | the object with light rings are black holes. Our theorem provides strong support for this assertion. \\

We establish that any stationary, axisymmetric, and asymptotically flat spacetime in $1+3$ dimensions with an ergoregion must have at least one light ring outside the ergoregion. Our proof does not assume anything about the nature of the central object except that it is rotating fast enough to develop an ergoregion. Then, the boundary conditions on the metric coefficients at the ergoregion and the asymptotic infinity are enough to affirm the presence of the light ring. So, our result actually proves the assumption of the existence of at least one light ring used in \cite{Cunha1}, at least for spacetimes with an ergoregion. Together with the results in \cite{Cunha2}, \cite{Cunha1} and \cite{Friedman,Cardoso:2014sna}, our theorem suggests that the observation of a light ring is strong evidence for the existence of black holes. \\

An important possible extension of our result could be to understand the situation for a star that is not compact enough to have an ergoregion. In principle, the size of a star can always be such that there is no light ring outside. However, let us consider the case when the star is quite compact and its radius is slightly bigger than that of the ergoregion of the external spacetime. Note, since this hypothetical ergoregion is now within the rotating body, it is not a part of the physical spacetime. Then, the existence of the ergoregion would have allowed us to impose a boundary condition $L(r_e, \theta) > R(r_e, \theta)$ at its location. We expect that very close to the surface of the star of radius $r_s \gtrsim r_e$, we also have $L(r_s, \theta) > R(r_s, \theta)$. This will immediately imply the existence of a light ring outside the surface of the star. Though this argument may not be fully rigorous, it seems to suggest that all sufficiently compact rotating stars with or without ergoregion possess at least one external light ring.\\

We also like to generalize our results to higher dimensions. In that case, we may need to assume the existence of more than one spacelike Killing vector. Also, even in four dimensions, a general stationary spacetime may not have any further symmetry. An extension of our proof for such a case requires an appropriate notion of light rings in the absence of axisymmetry.
\\

\section*{Acknowledgement}
We thank Pedro Cunha and Carlos Herdeiro for discussion and comments on a previous draft. We also thank the anonymous referee for useful suggestions and comments on a previous version of the manuscript. The research of RG is supported by the Prime Minister Research Fellowship (PMRF-192002-120), Government of India. The research of SS is supported by the Department of Science and Technology, Government of India under the SERB CRG Grant (CRG/2020/004562).


\begin{thebibliography}{100}



\section*{\bf{References}}   

\bibitem{Akiyama:2019cqa}
K.~Akiyama \textit{et al.} [Event Horizon Telescope],
Astrophys. J. \textbf{875}, no.1, L1 (2019)
[arXiv:1906.11238 [astro-ph.GA]].\\



\bibitem{Cunha2}
P.~Cunha, V.P. and C.~A.~R.~Herdeiro,
Phys. Rev. Lett. \textbf{124}, no.18, 181101 (2020)
doi:10.1103/PhysRevLett.124.181101
[arXiv:2003.06445 [gr-qc]].\\




\bibitem{Wei}
S.~W.~Wei,
Phys. Rev. D \textbf{102}, no.6, 064039 (2020)
doi:10.1103/PhysRevD.102.064039
[arXiv:2006.02112 [gr-qc]].\\



\bibitem{Cunha1}
P.~Cunha, V.P., E.~Berti and C.~A.~R.~Herdeiro,
Phys. Rev. Lett. \textbf{119}, no.25, 251102 (2017)
doi:10.1103/PhysRevLett.119.251102
[arXiv:1708.04211 [gr-qc]].\\

\bibitem{Guo:2020qwk}
M.~Guo and S.~Gao,
Phys. Rev. D \textbf{103}, no.10, 104031 (2021)
doi:10.1103/PhysRevD.103.104031
[arXiv:2011.02211 [gr-qc]].\\

\bibitem{Friedman}
J. L. Friedman, ``Ergosphere instability", Commun.Math.Phys. 63 (1978) 243-255; DOI: 10.1007/BF01196933.\\

\bibitem{Chirenti:2008pf}
C.~B.~M.~H.~Chirenti and L.~Rezzolla,
Phys. Rev. D \textbf{78}, 084011 (2008)
doi:10.1103/PhysRevD.78.084011
[arXiv:0808.4080 [gr-qc]].


\bibitem{Cardoso:2014sna}
V.~Cardoso, L.~C.~B.~Crispino, C.~F.~B.~Macedo, H.~Okawa and P.~Pani,
Phys. Rev. D \textbf{90}, no.4, 044069 (2014)
doi:10.1103/PhysRevD.90.044069
[arXiv:1406.5510 [gr-qc]].\\


\bibitem{Hawking}
 S. W. Hawking, Black holes in general relativity. Commun. Math. Phys. 25, 152-166 (1972).\\




\bibitem{Mishra:2019trb}
A.~K.~Mishra, S.~Chakraborty and S.~Sarkar,
Phys. Rev. D \textbf{99}, no.10, 104080 (2019)
doi:10.1103/PhysRevD.99.104080
[arXiv:1903.06376 [gr-qc]].



\end{thebibliography}
\end{document}